\documentclass[11pt,a4paper,reqno]{article}
\usepackage{amssymb}
\usepackage{mathrsfs}
\usepackage{graphicx}
\usepackage{flafter}
\usepackage{amsmath}
\usepackage{fancyhdr}
\usepackage{makeidx}
\usepackage{textcomp}
\usepackage{makeidx}

\begin{document}

\title{Basket Options Valuation for a Local Volatility 
Jump-Diffusion Model with the Asymptotic Expansion Method}
\author{Guoping Xu and Harry Zheng
\thanks{Department  of Mathematics,
Imperial College, London SW7 2AZ, UK.
Email: guoping.xu@citi.com and  h.zheng@imperial.ac.uk}\\
Imperial College London}
\date{}
\maketitle

\noindent
{\bf Abstract} \ 
In this paper we discuss the   
basket options valuation for a jump-diffusion model. 
The underlying asset prices follow some correlated local volatility 
diffusion processes with
systematic  jumps. We derive a forward partial integral differential equation
(PIDE) for general stochastic processes and use  
the asymptotic expansion method  to approximate 
 the conditional expectation of the stochastic variance associated with the basket value process.  
The numerical tests show that the suggested method is fast and accurate in comparison with the Monte Carlo and other methods in most cases.

\bigskip\noindent{\bf Keywords} \
Basket options pricing, local volatility jump-diffusion model, 
forward PIDE, asymptotic expansion.

\bigskip\noindent{\bf Mathematics Subject Classification (2000)} \ 
91B28, 60G99

\bigskip\noindent{\bf JEL Classification} \ G13

\bigskip\noindent{\bf Insurance Mathematics Classification} \ IM12, IM20

\section{Introduction}
A basket option is an exotic option whose payoff depends
on the value of a portfolio of assets. 
Basket options are in general difficult to price and hedge 
due to the lack of analytic characterization of the distribution
of the sum of correlated random variables. 
Monte Carlo simulation is often used to price basket options, which is 
simple, accurate, but time-consuming. There has been some 
extensive research recently for fast and accurate  pricing methods.

Most work in the literature assumes that 
underlying asset prices follow geometric Brownian motions.  
The basket value is then the sum of correlated lognormal variables. 
The main idea of the analytic approximation method is to find a simple random
variable to approximate the basket value and then to use it to get 
a closed form pricing formula. The approximate random variable is 
required to match some moments of the basket value.
Levy (1992) uses a lognormal variable to approximate the basket value with the matched first and second moments. The results are remarkably good but there is no error estimation. 
Curran (1994) introduces the idea of conditioning variable
and conditional moment matching. 
The option price is  decomposed into two parts: 
one can be calculated exactly and the other approximately 
by conditional moment matching method. 
Rogers and Shi (1995) derive the lower and upper bounds.
Vanmaele et al.~(2004) suggest a  moment matching comonotonic 
approximation for basket options.
See Lord (2006) for other methods and references.
Efforts have been made to extend to more  general asset price models. 
Albrecher and Predota (2004) discuss  
the NIG L\'{e}vy process,
and Flamouris and Giamouridis (2007) the Bernoulli jump diffusion model.

Xu and Zheng (2009) suggest a jump-diffusion model 
for underlying asset price processes. 
The innovative feature of that model is that, apart from correlated
Brownian motions, there are  two types of
Poisson jumps: a systematic jump that affects all asset prices and 
idiosyncratic jumps that only affect specific asset prices. Such a model
can characterize well the market-wide phenomenon and individual events. 
Xu and Zheng (2009) use the partial exact approximation (PEA) method
to find a closed form approximate solution which  is  guaranteed to lie between the lower and 
upper bounds.  The numerical tests show that the PEA method  
has superior performance in comparison with other methods such as 
the lower bound, reciprocal gamma, and lognormal approximations.
The limitation of 
the PEA method is that it depends crucially on the conditioning variable which is derived from the estimation of the basket value and the closed form solutions of individual asset prices. This may not be possible for general processes. For example, if individual asset prices follow some local volatility models, see Dupire (1994), then there are in general no closed form solutions and the PEA  method cannot be applied.

To price basket options for general asset price processes one may study 
directly the basket value and its associated stochastic processes
which may contain stochastic volatilities and/or stochastic jump intensities and sizes. Dupire (1994) and
Derman and Kani (1998) show that any diffusion model
with stochastic volatility can be replaced by a local volatility model without
changing the European option price and the 
marginal distribution of the underlying asset price
thanks to the uniqueness of the solution to the corresponding pricing equation,
a parabolic PDE. In fact, Gy\"ongy (1986) discovers 
the equivalence of a non-Markovian model with 
a Markovian model and proves that one-dimensional margins of any It\^o process can be matched
by a one-dimensional Markovian local volatility process, that is,
the value of the square of the local volatility is equal to the expectation of the square of the stochastic volatility conditional
on the final stock price being equal to the strike price. 
Without the uniqueness one cannot
claim  the equivalence of the stochastic and local volatility models
but can still derive the same pricing equation.
Bentata and Cont (2009) 
derive a discontinuous analogue of Gy\"ongy's result 
for semimartingale asset price processes.

The pricing equation for general asset price processes 
may contain coefficients expressed in terms of some conditional expectations.
It is in general a challenging task to compute these conditional expectations
as there is no closed
form solution to the related SDE. One then tries to find some good approximations.
 Antonov et al.~(2009)
use the Markovian projection onto a  displaced diffusion and 
Avellaneda et al.~(2002)  apply the steepest decent search with
Varadhan's formula, both methods require to
solve some minimization problems. Xu and Zheng
(2009) derive a closed form approximation to the conditional expectation with 
a weighted sum of the lower bound and the conditional second moment adjustments.

In this paper we discuss the European basket options pricing for a 
local volatility jump-diffusion model. The main idea is to 
reduce a multi-dimensional local volatility jump-diffusion 
model problem to a
one-dimensional stochastic volatility jump-diffusion model, 
then to  derive a forward PIDE for the basket options price
with an unknown conditional expectation, or local volatility function, and finally
to apply the asymptotic expansion method to  
approximate the local volatility function.
The main contributions of the paper
to the existing literature of the basket options pricing are 
the following: we propose a correlated 
local volatility jump-diffusion model for underlying
asset price processes and  derive a forward PIDE for  general
asset price 
processes with stochastic volatilities and stochastic jump compensators, which
may be used for other applications in pricing and calibration,
and we find the approximation of the conditional expectation
with  the asymptotic expansion method. 
Numerical tests show that the method discussed in the paper, the
  asymptotic expansion method,  performs very well for most
 cases in comparison with the Monte Carlo method  and the PEA
 method discussed in Xu and Zheng (2009).

The paper is organized as follows. Section 2 formulates
 the basket options pricing problem, reviews some pricing results on jump-diffusion
asset price models  and forward PIDEs.  Section
 3 discusses the computation of the conditional expectation and applies the
 asymptotic expansion method to approximate the local volatility function.
 Section 4  elaborates the numerical implementation and compares
  the numerical performance of different methods in pricing basket
options. Section 5 is the conclusion.
The appendix contains the outline of the derivation of a forward
PIDE for a general stochastic process.

\section{Forward PIDE for Basket options pricing}
Assume a portfolio is composed of $n$ assets and the risk-neutral 
asset prices $S_i$ satisfy the following SDEs:
\begin{eqnarray}\label{e2}
\frac{d S_i(t)}{S_i(t-)}=(r-\lambda m)dt+\sigma_i(t,S_i(t-))d
W_i(t)+(e^Y-1)d N(t)
\end{eqnarray}
where $W_i(t)$ are standard Brownian motions 
with correlation coefficients $\rho_{ij}$ 
between $W_i(t)$ and $W_j(t)$,  $(e^Y-1)d N(t)$ is a differential form of
a compound Poisson process $\sum_{k=1}^{N(t)} (e^{Y_k}-1)$ with
$N$ being  a Poisson process with 
intensity $\lambda$ and  $\{Y_k\}$ iid normal variables 
with mean $\eta$ and variance $\gamma^2$, $e^{Y_k}-1$ is the proportional
change of the asset prices at the $k$th jump of the Poisson process, 
$m=\mathbf{E}[e^{Y_1}-1]=e^{\eta+{1\over 2}\gamma^2}-1$, $\sigma_i(t,S)$
are bounded local volatility functions, $r$ is
constant risk-free interest rate, and $N$, $W$ and
$\{Y_k\}$ are  independent to each other.
The basket value at time $t$ is given by 
\begin{eqnarray} {S}(t)=\sum_{i=1}^{n} w_i S_i(t),\nonumber \end {eqnarray}
where $w_i$ are positive constant weights. The basket value follows the SDE
\begin{eqnarray}{\label{eq090728.1}}\frac{d S(t)}{S(t-)}=(r-\lambda m)dt+\sum_{i=1}^{n} w_i\sigma_i(t,S_i(t-))\frac{S_i(t-)}{S(t-)}d
W_i(t)+(e^Y-1)d N(t).
\end{eqnarray}
Define
\begin{eqnarray}V(t)^2:=\frac{1}{S(t)^2}\sum_{i,j=1}^{n} 
w_iw_j\sigma_i(t,S_i(t))\sigma_j(t,S_j(t))S_i(t)S_j(t)\rho_{ij}
\end{eqnarray}
then ({\ref{eq090728.1}}) becomes
\begin{eqnarray}{\label{eq090728.2}}\frac{d S(t)}{S(t-)}=(r-\lambda m)dt+V(t)
dW(t)+(e^Y-1)d N(t)
\end{eqnarray}
with the initial price $S(0)=\sum_{i=1}^{n}\omega_i S_i(0)$ and
a standard Brownian motion $W$.
Note that $V(t)$ is a stochastic volatility which depends on
individual asset prices, not just the basket price.
We have a stochastic volatility jump-diffusion asset price model for the
basket option problem. We next review some related pricing results for jump-diffusion
asset price processes.

Andersen and Andreasen (2000) model the risk-neutral asset price $S$ by 
a jump-diffusion process 
\begin{eqnarray}{\label{eq090308.1}}\frac{d S(t)}{S(t-)}=(r(t)-q(t)-\lambda(t)m(t))dt+\sigma(t,S(t-))d
W(t)+(e^{Y(t)}-1)d N(t)
\end{eqnarray}
where $W(t)$ is a Brownian motion, $N(t)$ a Poisson process with deterministic
intensity $\lambda(t)$, $Y(t)$ a sequence of independent
random variables  with time dependent density $\zeta(\cdot;t)$, $\sigma$
a bounded time and state dependent local volatility function, $m$ a
deterministic function given by $m(t)=\mathbf{E}[e^{Y(t)}-1]$, $r$ and $q$
deterministic risk-free interest rate and dividend yield, and $N$, $W$, 
$Y$ independent to each other.
Andersen and Andreasen (2000) derive a forward PIDE
 for the European call option price $C(T,K)$ at time 0 as
a function of maturity $T>0$ and exercise price $K\geq 0$:
\begin{eqnarray}{\label{eq090308.2}}
C_T(T,K)&=& -q(T)C(T,K) +(q(T)-r(T)+\lambda(T)m(T))KC_K(T,K)\nonumber\\
&& {}+\frac{1}{2}\sigma(T,K)^2K^2C_{KK}(T,K)\nonumber\\
&& {} + \lambda(T)\left(
\int^{\infty}_{-\infty}C(T,Ke^{-y})e^{y}
\zeta(y;T)dy-(1+m(T))C(T,K) \right)
\end{eqnarray}
with the initial condition $C(0,K)=(S(0)-K)^{+}$.
Andersen and Andreasen (2000) also discuss the stochastic volatility jump-diffusion model
and point out that the European call option price
satisfies the same PIDE (\ref{eq090308.2}) 
with the local volatility function $\sigma$ replaced by
\begin{eqnarray}{\label{eq090309.3}}
\sigma(T,K)^2&=&\mathbf{E}[V(T)^2|S(T)=K]
\end{eqnarray} 
where $V(t)$ is a stochastic volatility process.

Carr et al.~(2004)  generalize the asset price model of 
Andersen and Andreasen (2000) to a general 
local volatility and local L\'evy type model given by
\begin{eqnarray}{\label{eq090309.6}}\frac{d S(t)}{S(t-)}=(r(t)-q(t))dt+\sigma(t,S(t-))d
W(t)+\int_{-\infty}^{\infty}(e^x-1)[\mu(dx,dt)-\nu(dx,dt)]
\end{eqnarray}
where 
$\mu(dx,dt)$ is a random counting measure and 
 $\nu(x,t)$  a compensator 
with a form $\nu(dx,dt)=a(S(t),t)k(x)dxdt$, 
here $a(S,t)$ is  a deterministic local speed function
that  measures the speed
at which the L\'evy process is running at time $t$ and stock price $S$,
and $k(x)dx$ specifies the arrival rate
of jumps of size $x$ per unit of time, 
all other terms are the same as in equation (\ref{eq090308.1}). 
Carr et al. (2004) derive a  PIDE for the
European call option price $C(T,K)$ at time 0  as
\begin{eqnarray}C_T(T,K)&=&-q(T)C(T,K)+(q(T)-r(T))KC_{K}(T,K)
+\frac{1}{2}\sigma(T,K)^2K^2C_{KK}(T,K)\nonumber\\
{\label{eq090310.4}}&&+\int_{0}^{\infty}a(T,z)C_{zz}(T,z)z\psi_e\left(\ln\frac{K}{z}\right) dz
\end{eqnarray}
with the initial condition $C(0,K)=(S(0)-K)^{+}$,
where $\psi_e$ is the double-exponential tail of the L\'evy measure given
by  
$$\renewcommand{\arraystretch}{1.5}
\psi_e(y) = \left\{ \begin{array}{ll}
\int_{y}^{\infty}(e^x -e^{y})k(x)dx & \textrm{for $  y>0$}\\
\int_{-\infty}^{y}(e^y -e^{x})k(x)dx  & \textrm{for $  y<0$.}
\end{array} \right.
$$
Kindermann et al.~(2008) show the existence and uniqueness of the solution to the PIDE (\ref{eq090310.4}) under some continuity and uniform positive definiteness conditions. 
Carr and Wu (2009)  generalize the local volatility
asset price process (\ref{eq090309.6}) further to  a stochastic volatility asset price process
with a stochastic jump compensator $\nu(dx,dt)=\bar a(t)k(x)dx 
dt$ and  $\bar a(t)$ being the stochastic instantaneous variance. 
Carr and Wu (2009) use the model and the fast Fourier transform to value
stock options and credit default swaps in a joint framework. 
We can show that the European call option
price $C(T,K)$ satisfies the  PIDE (\ref{eq090310.4}) for a general stochastic
process $\bar a$ with the local volatility
function $\sigma$ given by  (\ref{eq090309.3}) and the local speed function
$a$ given by 
\begin{equation} \label{speedfunc}
a(T,z)= \mathbf{E}[\bar a(T)|S(T)=z]
\end{equation}
where $\mathbf{E}[\bar a(T)|S(T)]$ is the conditional expectation of $\bar
a(T)$ given $S(T)$. 
The derivation of the  PIDE (\ref{eq090310.4}) with
the local volatility function (\ref{eq090309.3}) and the local speed function
(\ref{speedfunc}) is given in the appendix.

\section{Approximation of Local Volatility Function}
For the basket value process (\ref{eq090728.2}) the corresponding 
local volatility function is given by
\begin{eqnarray*}\sigma(T,K)^2
&=&\frac{1}{K^2}\sum_{i,j=1}^{n} w_iw_j\rho_{ij}\mathbf{E}
[\hat\sigma_i(T,S_i(T))\hat\sigma_j(T,S_j(T))|S(T)=K]\quad\quad
\end{eqnarray*}
and $\hat\sigma_i(T,S_i(T))=\sigma_i(T,S_i(T))S_i(T)$. 
Piterbarg (2007)  uses the Taylor formula to approximate
$\hat\sigma_i(T,S_i(T))$ to the first order with respect to $S_i(T)$ 
at point $F_{i}\equiv S_i(0)e^{rT}$ to get 
$$ \hat\sigma_i(T,S_i(T)) \approx p_i+q_i(S_i(T)-F_i)$$
where $p_i=\hat\sigma_i(T,F_i)$ 
and $q_i={\partial\over \partial F_i}\hat\sigma_i(T,F_i)$.
We use the same first order approximation to get
\begin{eqnarray*}\hat\sigma_i(T,S_i(T))\hat\sigma_j(T,S_j(T))
&\approx&p_ip_j+p_j q_i(S_i(T)-F_i)+p_i q_j(S_j(T)-F_j).
\end{eqnarray*}
If we define $\hat \sigma(T,K)^2=\sigma(T,K)^2K^2$, then 
\begin{eqnarray}{\label{eq090805.1}}\hat \sigma(T,K)^2\approx\sum_{i,j=1}^{n} w_iw_j\rho_{ij}p_ip_j(1+\varphi_i(T,K)+\varphi_j(T,K))
\end{eqnarray}
where $\varphi_i(T,K)=\frac{q_i}{p_i}\mathbf{E}[S_i(T)-F_i|S(T)=K]$.

Without loss of generality we may assume $r=0$ 
(otherwise we can work on discounted asset price processes) so $F_i=S_i(0)$. To obtain analytical approximation to $\mathbf{E}[S_i(T)-S_i(0)|S(T)=K]$,
we use the asymptotic expansion approach related to small diffusion and small jump intensity and size, see Benhamou et al.~(2009). The perturbation and
its purpose are different in this paper. In Benhamou et al.~(2009) the authors
expand a parameterized process to the second order
and apply it directly to price European options. In this paper we use a 
different parameterized process and expand it to the first order to get the
analytic tractability and use it to approximate the conditional expectation of stochastic variance. In other words, we use the asymptotic expansion to
find the {\it unknown} 
local volatility function and then use it in the forward PIDE, 
while Benhamou et al.~(2009) use a different asymptotic
expansion to a process with a {\it known} local volatility function and then
find the options value directly.  
Assume $\epsilon\in [0,1]$ and define 
\begin{eqnarray*}d S_i^\epsilon(t)=-\lambda m^\epsilon S_i^\epsilon(t-) dt
+\epsilon\hat\sigma_i(t,S_i^\epsilon(t))d
W_i(t)+S_i^\epsilon(t-)(e^{\epsilon Y}-1)d N(t)
\end{eqnarray*}
with the initial condition $S_i^\epsilon(0)=S_i(0)$, 
where $m^\epsilon=\mathbf{E}[e^{\epsilon Y}-1]=e^{\epsilon\eta+\frac{1}{2}\epsilon^2\gamma^2}-1$.
 Note that $S_i^1(T)=S_i(T)$. If we define $S_{i,k}(t)=\frac{\partial^k S_i^\epsilon(t)}{\partial
 \epsilon^k}|_{}\epsilon=0$,
  then the first order asymptotic expansion around
 $\epsilon=0$ for $S_i^\epsilon(T)$ is
\begin{eqnarray} \label{dag} S_{i}^\epsilon(T)\approx S_{i,0}(T)+S_{i,1}(T)\epsilon.
\end{eqnarray}
We can find $S_{i,0}(T)$ and $S_{i,1}(T)$ as follows:
$S_{i,0}$ satisfies the equation $d S_{i,0}(t)=0$ with the initial condition
$S_{i,0}(0)=S_i(0)$, therefore, $S_{i,0}(t)\equiv S_i(0)$ for all $t$. 
$S_{i,1}$ satisfies the equation
\begin{equation} d S_{i,1}(t)
=-\lambda \eta S_{i}(0) dt+\hat\sigma_i(t,S_{i}(0))d
W_i(t)+S_{i}(0)Y dN(t) \label{e1}
\end{equation}
with the initial condition $S_{i,1}(0)=0$.
Here we have used the result $S_{i,0}(t)=S_i(0)$. Therefore, 
\begin{eqnarray}S_{i,1}(T)=-\lambda \eta S_{i}(0) T+\int_0^T\hat\sigma_i(t,S_{i}(0))d
W_i(t)+S_{i}(0)\sum_{l=1}^{N(T)}Y_l. 
\end{eqnarray}
The asset value $S_i(T)$ at time $T$ may be approximated by
\begin{eqnarray*}S_i(T)=S_{i}^1(T)\approx S_{i,0}(T)
+S_{i,1}(T)=S_i(0)+S_{i,1}(T)
\end{eqnarray*}
and the basket value by
\begin{eqnarray}{\label{eq091017.1}}S(T)
\approx S(0) + \sum_{i=1}^{n}\omega_i S_{i,1}(T) := S_c(T).
\end{eqnarray}
Conditional on $N(T)=k$, the variable $S_{i,1}(T)$, written as
$S_{i,1}(T,k)$, is a normal variable with mean 
$(-\lambda  T+k)\eta S_{i}(0)$ 
and variance $\int_{0}^{T}\hat\sigma_i^2(t,S_{i}(0))dt +k\gamma^2 S_{i}(0)^2$, and the variable $S_c(T)$, written as $S_c(T,k)$, is also a normal variable with mean
\begin{equation} \label{muc}
\mu_c(k) = (1-\lambda T\eta  +k \eta) S(0)
\end{equation}
and variance
\begin{equation} \label{sigmac}
\sigma_c(k)^2 = \sum_{i,j=1}^{n}\omega_i\omega_j \left[(\int_0^T\hat\sigma_i(t,S_{i}(0))\hat\sigma_j(t,S_{j}(0))dt)\rho_{ij}
+k \gamma^2S_{i}(0)S_{j}(0)\right].
\end{equation}
Therefore,
\begin{eqnarray*}\mathbf{E}[S_i(T)-S_i(0)|S(T)=K]
&\approx&\mathbf{E}[S_{i,1}(T)|S_c(T)=K]\\
&=&\sum_{k=0}^{\infty}P(N(T)=k)\mathbf{E}[S_{i,1}(T,k)|S_c(T,k)=K].
\end{eqnarray*}
Since $S_{i,1}(T,k)$ and $S_c(T,k)$ are normal variables, 
 we can find  $\mathbf{E}[S_{i,1}(T,k)|S_c(T,k)=K]$ exactly
 as 
\begin{eqnarray*}
\mathbf{E}[S_{i,1}(T,k)|S_c(T,k)=K]
=\mathbf{E}[S_{i,1}(T,k)] + 
\frac{C_{i}(k)}{\sigma_c(k)^2}(K-\mu_c(k))
\end{eqnarray*}
where $C_{i}(k)$ is the covariance of $S_{i,1}(T,k)$ and
$S_c(T,k)$, given by
\begin{eqnarray*}
C_{i}(k)
&=&\sum_{j=1}^{n}\omega_j \left[\rho_{ij}(\int_0^T\hat\sigma_i(t,S_{i}(0))\hat\sigma_j(t,S_{j}(0))dt)
+k \gamma^2S_{i}(0)S_{j}(0)\right].
\end{eqnarray*}
We obtain $\varphi_i(T,K)$ in ({\ref{eq090805.1}})  as
\begin{eqnarray*}
\varphi_i(T,K)=\frac{q_i}{p_i}\sum_{k=0}^{\infty}P(N(T)=k)
\frac{C_{i}(k)}{\sigma_c(k)^2}(K-(1-\lambda
T\eta + k\eta)S(0))\quad
\end{eqnarray*}
and $\hat \sigma(T,K)^2$ in ({\ref{eq090805.1}})  as
$$
\hat \sigma(T,K)^2= a(T) + b(T)K- c(T)S(0)
$$
where 
\begin{eqnarray*}
a(T) &=& \sum_{i,j=1}^{n} w_iw_j\rho_{ij}p_ip_j\\
b(T) &=& \sum_{i,j=1}^{n} \sum_{k=0}^{\infty}
\frac{P(N(T)=k)}{\sigma_c(k)^2}w_iw_j\rho_{ij}p_ip_j
\left(\frac{q_i}{p_i}C_{i}(k)+\frac{q_j}{p_j}C_{j}(k)\right)\\
c(T) &=& \sum_{i,j=1}^{n} \sum_{k=0}^{\infty}
\frac{P(N(T)=k)}{\sigma_c(k)^2}w_iw_j\rho_{ij}p_ip_j
\left(\frac{q_i}{p_i}C_{i}(k)+\frac{q_j}{p_j}C_{j}(k)\right)(1-\lambda
T\eta + k\eta).
\end{eqnarray*}
The European basket call option price $C(T,K)$ at time 0 satisfies the 
 PIDE (\ref{eq090308.2}), i.e., 
\begin{eqnarray} 
C_T(T,K)&=&\lambda m  KC_K(T,K)
+\frac{1}{2}\sigma(T,K)^2K^2C_{KK}(T,K) \nonumber \\
&& {} +\lambda
\int_{-\infty}^{\infty}C(T, Ke^{-y})
e^{y} \phi_{\eta,\gamma^2}(y)dy-\lambda (1+m) C(T,K)
\label{pide}
\end{eqnarray}
with the initial condition $C(0,K)=(S(0)-K)^{+}$, where
$\sigma(t,S)$ is a local volatility function given by
\begin{equation} \label{e3}
\sigma(t,S)=\frac{\sqrt{a(t)+b(t)S-c(t)S(0)}}{S}.
\end{equation}
and $\phi_{\eta,\gamma^2}$ is the density
function of a normal variable with mean $\eta$ and variance $\gamma^2$.

\section{Numerical Results}
In this section we do some numerical tests for the European basket  call options pricing
with the underlying asset price processes (\ref{e2}). 
We use three different methods to facilitate the comparison: 
the full Monte Carlo (MC), 
the  asymptotic expansion  (AE), and
the control variate (CV) method.  

The MC method  provides the benchmark results. 
We use the control variate technique to reduce the variance. 
In \eqref{eq091017.1} the basket value $S(T)$
is approximated by the first order
asymptotic expansion $S_c(T)$ which is used here
 as a control variate in MC simulation. 
The basket option price with the control variate $S_c(T)$ is given by 
\begin{eqnarray}
\mathbf{E}[(S_c(T)-K)^+]
&=&\sum_{k=0}^{\infty}P(N(T)=k)\mathbf{E}[(S_c(T,k)-K)^+]. \label{cv1}
\end{eqnarray}
Since $S_c(T,k)$ is a normal variable with mean $\mu_c(k)$ and 
variance $\sigma_c(k)^2$, see (\ref{muc}) and
(\ref{sigmac}), a trivial calculation shows that
\begin{equation}
\mathbf{E}[(S_c(T,k)-K)^+]
= \sigma_c(k)\varphi\left(\frac{K-\mu_c(k)}{\sigma_c(k)}\right)+
(\mu_c(k)-K)\Phi\left(\frac{-K+m_c(k)}{\sigma_c(k)}\right) \label{cv2}
\end{equation}
where $\phi$ is the density function of a standard normal variable
and $\Phi$ its 
cumulative distribution function.

The AE method is to solve the  PIDE (\ref{pide})
with the approximate local volatility function (\ref{e3}).
We find the numerical solution with the log transform  of variables and
the explicit-implicit finite difference  method of 
Cont and Voltchkova (2005). 

The CV method approximates the basket value $S(T)$ with
a tractable variable $S_c(T)$ and finds a closed form pricing formula 
 (\ref{cv1}) and (\ref{cv2}). This approach is essentially 
in  the same spirit as that of Benhamou et al.~(2009)
with the difference that 
 we only expand to the first order while Benhamou et al.~(2009) to the second order.

The following data are used in all numerical tests:
the number of assets in the basket $n=4$,
the portfolio weights  of each asset $w_i=0.25$ for $i=1,\ldots,n$, 
the correlation coefficients of Brownian motions 
$\rho_{ij}=0.3$ for $i,j=1,\ldots,n$, the initial asset
prices $S_i(0)=100$ for $i=1,\ldots,n$, the risk free interest rate $r=0$,
the dividend rate $q=0$, the exercise price $K=100$.

\medskip
\begin{center} \tt Table 1\end{center}
\medskip

Table 1 displays the numerical results of the European basket call option prices with the MC, AE, CV methods. The first column is the maturity
($T=1,3$), 
the second and the third the coefficients of local volatility functions
($\sigma(S)=\alpha S^{\beta-1}$ with $\alpha=01,0.2,0.5$ and $\beta=1,0.8,0.5$),
 the fourth the MC results with standard deviations in the brackets, 
 the fifth  
the AE results with relative percentage errors in comparison with the MC
results,  the sixth the CV results
with errors, columns four to six correspond to the case 
of jump intensity $\lambda=0.3$, and columns seven to nine 
are similarly defined for  
$\lambda=1$. The last row displays the average standard deviations of the
MC method and
the average errors of the AE and CV methods.
For normal variable $Y\sim N(\eta,\gamma^2)$
we  set $\eta=-0.08$ and $\gamma=0.35$. 
Whenever there is a jump event the jump size is 
relatively small
(about 2\% of the value lost).  The choice of $\eta, \gamma$ and
intensity $\lambda=0.3$ follow those of Benhamou et al.~(2009) where the authors claim that these parameters are not small, especially for the jump intensity $\lambda$ and the jump volatility $\gamma$. 

 It is clear  
that the overall performance of the AE method is excellent.
All relative errors are less than 0.5\% except for the four cases when
the local volatility function is $\sigma(S)=0.5$. This is the case corresponding
to the high volatility in the Black-Scholes setting and is irrelevant to
the maturity $T$ and jump intensity $\lambda$. This is the phenomenon also
reported by other researchers. The CV method is not satisfactory with average relative error about 7\%. We use the Matlab to do the computations. When $T=1$ we run 30,000 simulations  for each case and repeat 10 times to get
the average value, which is used as the Monte Carlo result. We choose the time step size
$1/512$ and state step size $1/1024$ for the explicit-implicit finite difference
method, it takes 40 seconds for the AE
method and more than 30 minutes for the MC method. When $T=3$ we run 
100,000 simulations for each case and repeat 10 times to get the average
Monte Carlo result and choose the same step sizes as those
for $T=1$, it takes 2
minutes for the AE method and more than 6 hours for the MC method. 
The AE 
method is much faster than the MC method while the accuracy is reasonable for most cases. 

\medskip
\begin{center} \tt Table 2\end{center}
\medskip

Table 2 is similar to Table 1 with the only difference that the mean 
of $Y$ is  $\eta=-0.3$. Whenever there is a jump event the jump size is 
relatively large
(about 21\% of the value lost). The performance of the AE method
is very similar to that in Table 1 with the average relative error
0.5\%, but the performance of the CV method becomes much worse 
with the average relative error 18\%. 
Since the CV method is similar in spirit to the method of Benhamou et al.~(2009) there is a possibility that large errors may appear when $\eta=-0.3$ in Benhamou et al.~(2009).

\medskip
\begin{center} \tt Table 3\end{center}
\medskip

Table 3 displays the results with three different methods: MC, AE, and
the partial exact approximation (PEA) method suggested in Xu and Zheng (2009) when the local volatility function is $\sigma(t,S)=0.2$
and the  variable  $Y$  is a constant $-\eta$. The purpose of the test is to see and compare  the performance of the AE and PEA methods.  The basic data are the same as those in Table 1 and 2. We perform
numerical tests for three constant jump sizes $m=e^{-\eta}-1$ with 
$\eta=0.25, 0.125, 0.0625$, which results in $m=-0.2212, -0.1175, -0.0606$, respectively. 
The first column is the jump intensity ($\lambda=0.3,1$), 
the second the jump size ($m=-0.2212, -0.1175, -0.0606$), 
the third maturity ($T=1,3$), 
the fourth the MC results, the fifth the PEA results with relative
errors compared with the MC results, the last the AE results with relative  errors. It is clear that both the PEA method and the 
AE method perform well with the relative error less than 1\%
 for all cases, and the former is  more accurate  
than the latter (the average relative error 0.1\% vs 0.4\%).

\medskip
\begin{center} \tt Table 4\end{center}
\medskip

Table 4 is similar to Table 3 with the difference that the local volatility function is changed to $\sigma(t,S)=0.5$. 
It is clear that the performance of the PEA method is much better than that of the AE method: the former has relative errors less than 1 percent for all cases while the latter has relative errors about 2 percent when $T=1$ and jumps to about 6 percent when $T=3$, irrespective to the jump intensities and sizes. We can reasonably say that the PEA method is a
better approximation method for the European basket call options pricing when the local volatility functions are of the Black-Scholes type. However, the AE method is much more flexible and can handle general local volatility functions (and stochastic volatilities) and general jump variables $e^{Y(t)}$, the two cases cannot be solved with the PEA method for the time being. 

\section{Conclusion}
In this paper we have discussed the European basket options pricing for 
a local volatility  jump-diffusion model. We have derived
 a forward PIDE for the European options price with general asset price processes. We have applied the asymptotic
expansion method to find the approximation of the conditional expectation of the correlated stochastic variance.
We have shown the excellent numerical performance of the AE method in comparison
with the Monte Carlo and other methods in most cases. The idea 
and methodology of the paper opens the way for other processes and refinements,
for example, we may get better approximation if we asymptotically expand
to the second order or we may introduce individual jump processes or different jump
sizes for the common jumps. We are currently working on these problems.

\bigbreak
\bigbreak\noindent
{\bf\Large Appendix}

\medskip\noindent
{\it Outline of the proof of the PIDE (\ref{eq090310.4})
with the local functions (\ref{eq090309.3}) and
(\ref{speedfunc})}.
According to Protter (2003), Theorem IV.68, 
\begin{eqnarray*}
(S(T)-K)^{+}
&=&(S(0)-K)^{+}+\int_{0}^{T}1_{[S(t-)>K]}d S(t)+\frac{1}{2}L_T^K\\
&&{}+\int_{0}^{T}\int_{-\infty}^{\infty}\left[1_{[S{(t-)}\leq K]}(e^x S(t-)-K)^{+}+1_{[S{(t-)}>K]}(K-e^x S(t-))^{+}\right]\mu(dx,dt)
\end{eqnarray*} 
where $L^K$ is a  local time at
$K$ of process $S$.  
Taking the expectation on both sides, using Fubini's theorem
and the martingale property, we have
\begin{eqnarray}
&&\mathbf{E}[(S(T)-K)^{+}]\nonumber\\
&=&(S(0)-K)^{+}+\int_{0}^{T}(r(t)-q(t))\mathbf{E}[1_{[S(t)>K]}S(t)]dt+\frac{1}{2}\mathbf{E}[L_T^K]\nonumber\\
{\label{eq090311.2}}&&{}+\int_{0}^{T}\mathbf{E}
\left[ \int_{-\infty}^{\infty}[1_{[S{(t)}\leq K]}(e^x S(t)-K)^{+}+1_{[S{(t)}>K]}(K-e^x S(t))^{+}]\bar a(t)k(x)dx \right]dt.
\end{eqnarray} 
We have replaced $S(t-)$ by $S(t)$ due to the time integral taken with
respect to the Lebesgue measure. 
Differentiating (\ref{eq090311.2}) with respect to T yields 
\begin{eqnarray}
\frac{\partial\mathbf{E}[(S(T)-K)^{+}]}{\partial T}
&=&(r(T)-q(T))\mathbf{E}[1_{[S(T)>K]}S(T)]+\frac{1}{2}\frac{\partial \mathbf{E}[L_T^K]}{\partial T}\nonumber\\
{\label{eq090311.3}}
&&{}+\mathbf{E}\left[\int_{-\infty}^{\infty}
L(T,K,x,S(T)) \bar a(T)k(x)dx\right]\quad\quad\quad
\end{eqnarray}
where
\begin{eqnarray*}
L(T,K,x,S(T)) = [1_{[S{(T)}\leq K]}(e^x S(T)-K)^{+}
+1_{[S{(T)}>K]}(K-e^x S(T))^{+}].
\end{eqnarray*}
Since the European call option price at time 0 with maturity $T$ and exercise price $K$ is given by
\begin{eqnarray}{\label{eq090309.1}}C(T,K)=e^{-\int_{0}^{T}r(t)dt}\mathbf{E}[(S(T)-K)^{+}].
\end{eqnarray}
we have (Klebaner (2002))
\begin{eqnarray}{\label{eq090308.9}}\mathbf{E}[1_{[S(T)>K]}]
=1-F_{S(T)}(K)=-\frac{\partial C(T,K)}{\partial
K}e^{\int_{0}^{T}r(t)dt}
\end{eqnarray}
where $F_{S(T)}$ is  the cumulative distribution function
of $S(T)$, and 
\begin{eqnarray}{\label{eq090914.1}}\frac{d F_{S(T)}(K) }{d
K}=e^{\int_{0}^{T}r(t)dt}C_{KK}(T,K).
\end{eqnarray}
Note that the above equation and derivatives 
are defined in the sense of distribution. 
If $S(T)$ admits a continuous probability density function
then $C(T,K)$ is twice
continuously differentiable and (\ref{eq090914.1}) holds in the classical sense. Since 
\begin{eqnarray*}\mathbf{E}[(S(T)-K)^{+}]=\mathbf{E}[1_{[S(T)>K]}S(T)]-K
\mathbf{E}[1_{[S(T)>K]}]
\end{eqnarray*} 
we can combine (\ref{eq090309.1}) with {\eqref{eq090308.9}} to yield
\begin{eqnarray*}
\mathbf{E}[1_{[S(T)>K]}S(T)]=e^{\int_{0}^{T}r(t)dt}C(T,K)-
K e^{\int_{0}^{T}r(t)dt}\frac{\partial C(T,K)}{\partial K}.
\end{eqnarray*} 
We also clearly have
\begin{eqnarray*}
\frac{\partial\mathbf{E}[(S(T)-K)^{+}]}{\partial T}
&=&{\partial \over\partial T}C(T,K)e^{\int_{0}^{T}r(t)dt}
+C(T,K)e^{\int_{0}^{T}r(t)dt}r(T)
\end{eqnarray*} 
Following the same proof as in Klebaner (2002), Theorem 4, we can show that \begin{eqnarray}{\label{eq090308.11}}
\frac{\partial \mathbf{E}[L_T^K]}{\partial T}
= \mathbf{E}[V(T)^2 K^2|S(T)=K] e^{\int_{0}^{T}r(t)dt}C_{KK}(T,K).
\end{eqnarray} 
The equation (\ref{eq090308.11}) and derivatives are defined
in the sense of distribution.
Klebaner (2002) proves (\ref{eq090308.11}) for continuous semimartingale asset price process, it also works for the case here thanks to the property of the local time, that it,
\begin{eqnarray*}\int_{-\infty}^{\infty}g(K)L_T^K d K
=\int_{0}^{T}g(S(t-))d\langle
S^c \rangle_t
\end{eqnarray*}
for all positive bounded functions $g$, where  
$\langle S^c \rangle_v$ is the quadratic variation of the continuous part of the process $S$. Everything then proceeds exactly the same. 
We now estimate the last term in (\ref{eq090311.3}). 
Using Fubini's theorem 
and the tower property, also noting (\ref{speedfunc}) and 
(\ref{eq090914.1}),  yield
\begin{eqnarray*}
&&\mathbf{E}\int_{-\infty}^{\infty}\left[L(T,K,x,S(T))\bar
a(T)\right]k(x)dx \\
&=&\int_{-\infty}^{\infty}\mathbf{E}\left[\mathbf{E}[L(T,K,x,S(T))\bar
a(T)|S(T)]\right]k(x)dx\\
&=&\int_{-\infty}^{\infty}\mathbf{E}\left[L(T,K,x,S(T))
a(T,S(T))\right]k(x)dx\\
&=& \int_{-\infty}^{\infty}\int_{0}^{\infty} L(T,K,x,z)a(T,z) dF_{S(T)}(z)k(x)dx\\
&=& \int_{-\infty}^{\infty}\int_{0}^{\infty} L(T,K,x,z)a(T,z) e^{\int_0^T r(t)dt} C_{zz}(T,z)dz k(x)dx\\
&=&e^{\int_0^T r(t)dt}
\int_{0}^{\infty}a(T,z)C_{zz}(T,z)\left(
\int_{-\infty}^{\infty}L(T,K,x,z)k(x)dx\right)dz\\
&=& e^{\int_0^T r(t)dt}
\int_{0}^{\infty}a(T,z)C_{zz}(T,z)z\psi_e\left(\ln\frac{K}{z}\right) dz
\end{eqnarray*} 
where $\psi_e$ is the double-exponential tail of the L\'evy measure $k$. The last equality follows exactly Carr et al. (2004).
Substituting everything into (\ref{eq090311.3}) and simplifying the expression we then get the PIDE (\ref{eq090310.4})
with local functions (\ref{eq090309.3}) and
(\ref{speedfunc}).

\newpage

\begin{table}
\begin{center}
{\small
\renewcommand{\arraystretch}{1.2}
\begin{tabular}{|c|r|r|rrr|rrr|}
\hline
\multicolumn{3}{|c|}{$\lambda$}
&  \multicolumn{3}{c|}{0.3}& \multicolumn{3}{c|}{1} \\ \hline
$T$& $\alpha$ & $\beta$& MC (stdev) &AE (err\%) & CV (err\%)& MC  (stdev) & AE (err\%) &CV (err\%)\\ \hline
1&0.1 & 1&5.91 (0.03) &5.91 (0.0)&6.14 (3.9)&11.86  (0.05)&11.83 (0.3)&12.7 (7.1)\\ 
&0.2&&8.14 (0.02)&8.13 (0.1) &8.31 (2.1)&13.25 (0.06) &13.24 (0.1)&13.85 (4.5)\\
&0.5&&15.50 (0.04)&15.18 (2.1)&15.52 (0.1)&18.89  (0.05)&18.60 (1.5)&19.25 (1.9)\\ \cline{2-9}
&0.1&0.8&4.64 (0.02)&4.64 (0.0)&5.03 (8.4)&11.13 (0.03)&11.16 (0.3)&12.43 (11.7)\\ 
&0.2&&5.47 (0.02)&5.47 (0.0)& 5.74 (4.9) &11.60 (0.06)&11.58 (0.2)&12.56 (8.3)\\
&0.5&&8.11 (0.03)&8.11 (0.0)&8.29 (2.2)&13.25 (0.04)&13.23 (0.2)& 13.84 (4.5)\\ \cline{2-9}
&0.1&0.5&4.06 (0.02)&4.08 (0.5)&4.81 (18.5)&10.96 (0.03)&10.98 (0.2)&12.40 (13.1) \\
&0.2&&4.24 (0.01)&4.25 (0.2)&4.83 (13.9)&11.00 (0.02)&11.01 (0.1)&12.41 (12.8)\\
&0.5&&4.85 (0.01)&4.85 (0.0)&5.19 (7.0)&11.24 (0.04)&11.26 (0.2)&12.44 (10.7)\\ \hline
3&0.1&1&12.18 (0.02)&12.16 (0.2)&12.75 (4.7)&22.94 (0.11)&22.99 (0.2)&24.36 (6.2)\\
&0.2&&15.25 (0.03)&15.14 (0.7)&15.65 (2.6)&24.49 (0.13)&24.45 (0.2) &25.78 (5.3)\\
&0.5&&27.23 (0.09)&25.64 (5.8)&27.21 (0.1)&33.03 (0.14)&31.55 (4.5) &34.02 (3.0)\\ \cline{2-9}
&0.1&0.8&10.69 (0.02)&10.68 (0.1)&11.65 (9.0) &22.45 (0.10)&22.51 (0.3) & 23.97 (6.8)\\
&0.2&&11.64 (0.02)&11.62 (0.2)&12.29 (5.6)&22.81 (0.10)&22.79 (0.1)& 24.19 (6.1)\\
&0.5&&15.19 (0.03)&15.14 (0.3)&15.62 (2.8)&24.51 (0.07)&24.48 (0.1)& 25.76 (5.1)\\ \cline{2-9}
&0.1&0.5&10.16 (0.02)&10.16 (0.0)&11.53 (13.5)&22.35 (0.04)&22.43 (0.4)& 23.90 (6.9)\\ 
&0.2&&10.29 (0.02)&10.29 (0.0)&11.55 (12.2)&22.43 (0.07)&22.44 (0.0)&23.91 (6.6)\\
&0.5&&10.92 (0.02)&10.91 (0.1)&11.77 (7.8)&22.51 (0.10)&22.57 (0.3) &24.01 (6.7)\\ \hline
\multicolumn{3}{|c|}{Average} & (0.03)
& (0.6) & (6.6) & (0.07)&(0.5) & (7.1)\\ \hline
\end{tabular}
\caption{\label{table1} 
 The comparison of  European basket call option prices with 
 the Monte Carlo (MC),  the
asymptotic expansion (AE), and the
control variate (CV) methods.
The  asset price processes are modelled by SDE (\ref{e2}).
The table displays results 
with different maturities $T$, local volatility functions 
$\sigma_i(t,S)=\alpha S^{\beta-1}$,
and jump intensities $\lambda$.
The numbers inside brackets in the MC columns are the standard deviations
and those in the AE and CV columns are the relative percentage errors in comparison with the MC results. The data used are: number of assets $n=4$,
weights $w_i=0.25$, correlation of Brownian motions $\rho_{ij}=0.3$,
initial asset prices $S_i(0)=100$, interest rate $r=0$, 
exercise price $K=100$, normal variable  
$Y\sim N(\eta,\gamma^2)$ with $\eta=-0.08$ and $\gamma=0.35$. }
}
\end{center}
\end{table}

\newpage
\begin{table}
\begin{center}
{\small
\renewcommand{\arraystretch}{1.2}
\begin{tabular}{|c|r|r|rrr|rrr|}
\hline
\multicolumn{3}{|c|}{$\lambda$}
&  \multicolumn{3}{c|}{0.3}& \multicolumn{3}{c|}{1} \\ \hline
$T$& $\alpha$ & $\beta$& MC (stdev) &AE (err\%) & CV (err\%)& MC  (stdev) & AE (err\%) &CV (err\%)\\ \hline
1&0.1&1&6.99 (0.02)&7.00 (0.1)&8.4 (20.2) &15.23 (0.03) &15.28 (0.3)&18.20 (20.0) \\ 
&0.2&&8.84 (0.01) &8.84 (0.0)&9.91 (12.1)&15.76 (0.05) &15.79 (0.2)&18.62 (18.2)\\
&0.5&&15.89 (0.02) &15.62(1.7)&16.51 (3.9)&20.24 (0.04) &20.02 (1.1)& 22.34 (10.4)\\ \cline{2-9}
&0.1&0.8&6.45 (0.01) &6.45 (0.0)&8.13 (26.1)&15.14 (0.03)&15.17 (0.2)& 18.09 (19.5)\\ 
&0.2&&6.72 (0.01) &6.73 (0.2)&8.24 (22.6)&15.20 (0.04) &15.23 (0.2)& 18.15 (19.4)\\
&0.5&&8.83 (0.02)&8.83 (0.0)&9.89 (12.0)&15.75 (0.02)&15.79 (0.3)&18.62 (18.2)\\ \cline{2-9}
&0.1&0.5&6.43 (0.01) &6.44 (0.2)&8.12 (26.3)&15.11 (0.03)&15.15 (0.3) &18.07 (19.6)\\
&0.2&&6.44 (0.01) &6.44 (0.0)& 8.12 (26.1)&15.12 (0.03)&15.16 (0.3)&18.07 (19.5)\\
&0.5&&6.49 (0.01) &6.49 (0.0)&8.14 (25.4)&15.15 (0.05)&15.19 (0.3)&18.10 (19.5)\\ \hline
3&0.1&1&14.70 (0.02) &14.71 (0.1)&17.46 (18.8)&27.00 (0.07)&27.03 (0.1)& 31.92 (18.2)\\
&0.2&&16.85 (0.03)&16.79 (0.4)&19.11 (13.4)&28.08 (0.06)&28.04 (0.1)&32.93 (17.3)\\
&0.5&&27.99 (0.04) &26.51 (5.3)&29.19 (4.3)&35.31 (0.06)&33.92 (3.9)& 39.44 (11.7)\\ \cline{2-9}
&0.1&0.8&14.27 (0.02) &14.29 (0.1)&17.12 (20.0)&26.64 (0.07)&26.74 (0.4)& 31.63 (18.7)\\
&0.2&&14.48 (0.02) &14.49 (0.1)&17.30 (19.5)&26.82 (0.07)&26.91 (0.3)&31.79 (18.5)\\
&0.5&&16.81 (0.02)&16.80 (0.1)&19.09 (13.6)&28.07 (0.06)&28.09 (0.1)&32.92 (17.3)\\ \cline{2-9}
&0.1&0.5&14.22 (0.01)&14.23 (0.1)& 17.07 (20.0)&26.58 (0.04) &26.68 (0.4)&31.58 (18.8)\\ 
&0.2&&14.23 (0.01)&14.25 (0.1)&17.07 (20.0)&26.62 (0.05)&26.69 (0.3)&31.59 (18.7)\\
&0.5&&14.31 (0.02)&14.32 (0.1)&17.15 (20.0)&26.71 (0.06)&26.78 (0.3)&31.66 (18.5)\\ \hline
\multicolumn{3}{|c|}{Average} & (0.02)
& (0.5) & (18.0) & (0.05) &(0.5) & (17.9)\\ \hline
\end{tabular}
\caption{\label{table1} 
 The comparison of  European basket call option prices with 
 the Monte Carlo (MC),  the
asymptotic expansion (AE), and the
control variate (CV) methods.
The  asset price processes are modelled by SDE (\ref{e2}).
The table displays results 
with different maturities $T$, local volatility functions 
$\sigma_i(t,S)=\alpha S^{\beta-1}$,
and jump intensities $\lambda$.
The numbers inside brackets in the MC columns are the standard deviations
and those in the AE and CV columns are the relative percentage errors in comparison with the MC results. The data used are: number of assets $n=4$,
weights $w_i=0.25$, correlation of Brownian motions $\rho_{ij}=0.3$,
initial asset prices $S_i(0)=100$, interest rate $r=0$, 
exercise price $K=100$, normal variable  
$Y\sim N(\eta,\gamma^2)$ with $\eta=-0.3$ and $\gamma=0.35$. }
}
\end{center}
\end{table}

\newpage
\begin{table}
\begin{center}
{\small
\renewcommand{\arraystretch}{1.2}
\begin{tabular}{|c|c|c|rrr|}
\hline
$\lambda$& $m$& $T$ & MC (stdev)  & PEA (err\%) & AE (err\%) \\ \hline
0.3&-0.2212&1&7.35 (0.01)&7.35 (0.0) &7.35 (0.0) \\
\cline{3-6}
&&3&12.93 (0.01) &12.92 (0.1)&12.85 (0.6)\\
\cline{2-6}
&-0.1175&1&6.08 (0.01) &6.08 (0.0)&6.07 (0.2)\\
\cline{3-6}
&& 3&10.57 (0.01) &10.56 (0.1)& 10.49 (0.8)\\
\cline{2-6}
& -0.0606 & 1&5.66 (0.01)&5.66 (0.0)&5.65 (0.2)\\
\cline{3-6}
&& 3&9.83 (0.01)&9.82 (0.1)& 9.74 (0.9)\\
\cline{1-6}
1&-0.2212&1&10.78 (0.01) &10.77 (0.1) &10.78 (0.0) \\
\cline{3-6}
&&3&18.64 (0.01) &18.63 (0.1)&18.57 (0.4)\\
\cline{2-6}
&-0.1175&1&7.28 (0.01)&7.28 (0.0)&7.28 (0.0)\\
\cline{3-6}
&& 3&12.65 (0.01)&12.64 (0.1)& 12.58 (0.6)\\
\cline{2-6}
& -0.0606 & 1&6.02 (0.01)&6.02 (0.0)&6.01 (0.2)\\
\cline{3-6}
&& 3&10.45 (0.01) &10.43 (0.2)& 10.37 (0.8) \\ \hline
\multicolumn{3}{|c|}{Average} & (0.01)
& (0.1) & (0.4) \\ \hline
\end{tabular}
\caption{\label{table3} 
The comparison of the Monte Carlo (MC), the
partial exact approximation (PEA), and the asymptotic expansion (AE) methods. The asset price processes are modelled by 
SDE (\ref{e2}). The table displays the results with different jump intensities
$\lambda$, jump sizes $m$, and maturities $T$. 
The numbers inside brackets in the MC columns are the standard deviations
and those in the PEA and AE columns are the relative percentage errors in comparison with the MC results.
The data used are: number of assets $n=4$,
weights $w_i=0.25$, correlation of Brownian motions $\rho_{ij}=0.3$,
initial asset prices $S_i(0)=100$, interest rate $r=0$, and exercise price $K=100$, local volatility function 
 $\sigma_i(t,S)=0.2$, and jump variable $Y$ a constant.}
}
\end{center}
\end{table}

\newpage
\begin{table}
\begin{center}
{\small
\renewcommand{\arraystretch}{1.2}
\begin{tabular}{|c|c|c|rrr|}
\hline
$\lambda$& $m$& $T$ & MC (stdev)  & PEA (err\%) & AE (err\%) \\ \hline
0.3&-0.2212&1&14.71 (0.01) &14.66 (0.3) &14.42 (2.0) \\
\cline{3-6}
&&3&25.69 (0.04) &25.44 (1.0)&24.14 (6.0)\\
\cline{2-6}
&-0.1175&1&14.08 (0.01)&14.03 (0.4)&13.79 (2.1)\\
\cline{3-6}
&& 3&24.61 (0.03)&24.39 (0.9)& 23.07 (6.3)\\
\cline{2-6}
& -0.0606 & 1&13.90 (0.01)&13.85 (0.4)&13.61 (2.1)\\
\cline{3-6}
&& 3&24.32 (0.04) &24.11 (0.9)& 22.77 (6.4)\\
\cline{1-6}
1&-0.2212&1&16.60 (0.01) &16.55 (0.3) &16.32 (1.7) \\
\cline{3-6}
&&3&28.80 (0.04) &28.55 (0.9)&27.28 (5.3)\\
\cline{2-6}
&-0.1175&1&14.64 (0.01)&14.59 (0.3)&14.35 (2.0)\\
\cline{3-6}
&& 3&25.52 (0.05) &25.28 (0.9)& 24.00 (6.0)\\
\cline{2-6}
& -0.0606 & 1&14.05 (0.01)&14.00 (0.4)&13.76 (2.1)\\
\cline{3-6}
&& 3&24.55 (0.04)&24.33 (0.9)& 23.02 (6.2) \\ \hline
\multicolumn{3}{|c|}{Average} &  (0.03)
& (0.6) & (4.0) \\ \hline
\end{tabular}
\caption{\label{table4} 
The comparison of the Monte Carlo (MC), the
partial exact approximation (PEA), and the asymptotic expansion (AE) methods. The asset price processes are modelled by 
SDE (\ref{e2}). The table displays the results with different jump intensities
$\lambda$, jump sizes $m$, and maturities $T$. 
The numbers inside brackets in the MC columns are the standard deviations
and those in the PEA and AE columns are the relative percentage errors in comparison with the MC results.
The data used are: number of assets $n=4$,
weights $w_i=0.25$, correlation of Brownian motions $\rho_{ij}=0.3$,
initial asset prices $S_i(0)=100$, interest rate $r=0$, and exercise price $K=100$, local volatility function 
 $\sigma_i(t,S)=0.5$, and jump variable $Y$ a constant.}
}
\end{center}
\end{table}

\end{document}